\documentstyle[preprint,tighten,aps]{revtex}
\begin{document}
\draft
\title{\bf Critical behavior of weakly-disordered anisotropic systems in two
dimensions }
\author{ Giancarlo Jug }
\address{ INFM--Istituto di Scienze Matematiche, Fisiche e Chimiche \\
Universit\`{a} di Milano a Como, Via Lucini 3, 22100 Como (Italy) \\ and
INFN--Sezione di Pavia, 27100 Pavia (Italy) }
\author{ Boris N. Shalaev }
\address{ A.F.Ioffe Physical \& Technical Institute, Russian Academy of
Sciences,\\ 194021 St.Petersburg (Russia) \cite{BNS},\\ and International
School for Advanced Study, Via Beirut 4, 34014, Trieste (Italy) \\ and
INFN--Sezione di Pavia, 27100 Pavia (Italy) }
\maketitle
\begin{abstract} \noindent The critical behavior of two-dimensional (2D)
anisotropic systems with weak quenched disorder described by the so-called
generalized Ashkin-Teller model (GATM) is studied. In the critical region this
model is shown to be described by a multifermion field theory similar to the
Gross-Neveu model with a few independent quartic coupling constants.
Renormalization group calculations are used to obtain the temperature
dependence near the critical point of some thermodynamic quantities and the
large distance behavior of the two-spin correlation function. The equation of
state at criticality is also obtained in this framework.  We find  that random
models described by the GATM belong to the same universality class as that of
the two-dimensional Ising model.  The critical exponent $\nu$ of the
correlation length for the 3- and 4-state random-bond Potts models is also
calculated in a 3-loop approximation. We show that this exponent is given by an
apparently convergent series in $\epsilon=c-\frac{1}{2}$ (with $c$ the central
charge of the Potts model) and that the numerical values of $\nu$ are very
close to that of the 2D Ising model. This work therefore supports the
conjecture (valid only approximately for the 3- and 4-state Potts models) of a
superuniversality for the 2D disordered models with discrete symmetries.
\end{abstract}

\pacs{PACS numbers: 05.50.+q, 05.70.Jk, 75.10.Hk, 75.40.Cx } \vfill \newpage

\section{ INTRODUCTION } \renewcommand{\theequation}{1.\arabic{equation}}
\setcounter{equation}{0}

The critical properties of two-dimensional random spin systems have been
extensively studied in the last few years \cite{DD,Shalaev2,Vik}.
Two-dimensional (2D) systems are particularly interesting due to a variety of
reasons. Firstly, there are numerous examples of layered crystals undergoing
continuous antiferromagnetic and structural phase transitions \cite{hrk,hiik}.
More recently, 2D and quasi-2D crystals have begun to be fabricated and studied
thanks to advances in deposition techniques, with an enormous increase in the
variety of physical phenomena to be investigated \cite{lnp}. Perfect crystals,
however, are the exception rather than the rule, with quenched disorder always
existing in different degrees. Even weak disorder may drastically affect the
critical behavior, according to the celebrated Harris criterion \cite{harr}.
Secondly, the conventional field-theoretic renormalization group (RG) approach
based on the standard $\phi^4$ theory in $(4-\epsilon)$-dimensions, and as
applied to study properties of disordered systems by Harris and Lubensky
\cite{HL} and Khmelnitskii \cite{KDE}, does not work in 2D due to the hard
restriction $\epsilon {\ll}1$. Similar considerations apply to the
$(2+\epsilon)$ low-temperature RG approach.  Thirdly, from a theoretical point
of view a most challenging problem is to establish the relationship between
random models and the corresponding conformal field theory (CFT) describing
these at criticality.

Some early exact results concerning the 2D random-bond Ising model (IM) with a
special type of disorder (where only the vertical bonds are allowed to acquire
random values, while the horizontal bond couplings are fixed) have been
obtained by McCoy and Wu \cite{McW}.  This type of 1D quenched disorder without
frustration was shown to smooth out the logarithmic singularity of the specific
heat; the frustrated case was considered by Shankar and Murthy \cite{ShM}.
Many years ago Dotsenko and Dotsenko \cite{DD} initiated some considerable
progress in the study of 2D random bond IMs by exploiting the remarkable
equivalence between this problem and the N=0 Gross-Neveu model. For weak
dilution the new temperature dependence of the specific heat was found to
become $C{\sim}\ln\ln\tau$,  $\tau=\frac{T-T_{c}}{T_{c}}$ being the
reduced      deviation from the critical temperature $T_c$. However, their
results concerning the two-spin correlation function at the critical point were
later reconsidered by Shalaev \cite{Shalaev1}, Shankar \cite{Sh1}, and Ludwig
\cite{Lud1}. By using the RG approach as well as the bosonization technique
these authors showed that the large-distance behavior  of this function at
criticality was the very same as in the pure case. Some convincing  arguments
in favor of the critical behavior  of the 2D IM with impurities as governed by
the pure IM fixed point had been given earlier by Jug \cite{Jug}. Recently, a
good number of papers devoted to Monte-Carlo simulations of the critical
behavior of the random Ising model have been published \cite{WS}.  Most
Monte-Carlo data are in good agreement with analytical results obtained in
\cite{Shalaev1,Sh1,Lud1}.  It should be mentioned however, that these
analytical results have been obtained by employing the replica method.  This,
on the one hand, is known to give reliable results only in the framework of
perturbation theory. On the other hand, the mathematical legitimacy of the
replica trick has not yet been established. Moreover, replicas (though being
very useful and convenient) appear          not to capture the essentials of
nonperturbative effects in the close vicinity of the phase transition point
(Griffiths phase) (see, for instance, \cite{KZ}).  The study of nonperturbative
effects in the critical properties of random systems is however beyond the
scope of the present paper.

Here, we point out that there is some scope to extend the previous analysis for
the 2D random  IM to other discrete-symmetry systems. A very interesting
problem consists in considering minimal CFT models with $c <1$ as perturbed by
randomness. These models comprise the 3- and 4-state Potts systems as
particular cases, and these have interesting applications to real 2D crystals
\cite{lnp}.  Because the critical exponent $\alpha$ is positive for all these
models the critical behavior is governed by a random fixed point in agreement
with the Harris criterion. Some years ago Ludwig \cite{Lud2,Lud3} and Ludwig
and Cardy \cite{LudC} made an attempt to calculate perturbatively the critical
exponents of the random 3-state Potts model. Their approach was essentially
based on the powerful CFT technique. More recently, Dotsenko, Pujol and Picco
\cite{DPP} obtained the critical exponents for the dilute 3-state Potts model
in a two-loop approximation by exploiting the Coulomb Gas representation for
the correlation functions and a special kind of $\epsilon$-regularization,
where $\epsilon$ stands here for the difference between the pure system's
central charge value and the conformal anomaly for the pure 2D IM
($\frac{1}{2}$). Dotsenko, Dotsenko, Picco and Pujol \cite{DDPP} have also
found the new universality class of the critical behavior as corresponding to
the broken replica symmetry proposed by Harris, Dotsenko, Stinchcombe and
Sherrington \cite{HDSS}.

Another interesting possibility is to study critical phenomena in 2D dilute
anisotropic systems with  many-component order parameters. The analysis of the
critical behavior of such systems in $(4-\epsilon )$-dimensions has been
developed in great detail many years ago \cite{com1}, but cannot be directly
applied to the 2D case. Therefore it would be interesting and important
to consider studying  these 2D models. This is the main goal of our paper. The
key ingredient of our treatment is a fermionization trick first suggested by
Shankar \cite{Sh2} for the N-color Ashkin-Teller model (see also
\cite{Gold,Shalaev3}).  This method is quite general and may be extended to
other systems.  The initial Landau Hamiltonian as written in terms of scalar
fields can be shown to map onto a multifermion field theory of the Gross-Neveu
type with a few independent quartic couplings. This transformation can be done
for Hamiltonians containing only even powers of each order parameter component,
the fourth-order term being an invariant of the hypercubic symmetry group (this
is the so-called generalized Ashkin-Teller model (GATM)).

The work presented in this paper is organized as follows. In Section II we
consider in brief the critical behavior of the weakly-disordered 2D Ising model
with random bonds, this being the central theme of this research field. The
tranfer matrix formalism is set up and the corresponding equations are written
down. The computation of the two-spin correlation function for pure and random
models at criticality is also reviewed. In Section III we give a description of
the fermionization trick allowing us to study the critical behavior of the pure
N-color Ashkin-Teller model. In Section IV the critical properties of two
interacting N- and M-color quenched disordered Ashkin-Teller models are
studied. The RG method is used to obtain the exact temperature dependence of
the correlation length, specific heat, susceptibility and spontaneous
magnetization near criticality, as well as the two-point spin-correlation
function and the equation of state at  the critical point. In Section V,
exploiting the approach of Dotsenko, Picco and Pujol and of Ludwig we compute
the critical exponent of the correlation length in a 3-loop approximation for
the weakly-disordered minimal models of CFT, in particular for the 3- and
4-state Potts models with random bonds. We find that while for the GATM the
introduction of disorder leads to critical behavior as characterised by the
random-bond IM fixed point, for the minimal models of CFT this Ising behavior,
conjectured by a number of authors recently for the 2D Potts models
\cite{conj}, is actually only approximate. The accuracy with which the Ising
values of the exponents is observed, however, justifies the use of the term
``IM superuniversality'' for all these models, when disordered.  Section V
contains a discussion and some concluding remarks.

\section{ THE TWO-DIMENSIONAL ISING MODEL WITH RANDOM BONDS }
\renewcommand{\theequation}{2.\arabic{equation}} \setcounter{equation}{0}

\subsection{ Transfer Matrix, Effective Action and RG for Thermodynamic
Functions }

We begin with the classical Hamiltonian of the 2D Ising model with random bonds
defined on a square lattice with periodic boundary conditions:

\begin{equation} H=-\sum_{i=1,j=1}^{N}[J_{1}(i,j)s_{ij}s_{ij+1}+J_{2}(i,j)
s_{ij}s_{i+1j}] \label{a} \end{equation}

\noindent where $i,j$ label sites of the square lattice, $s_{ij}=\pm1$ are spin
variables, $J_{1}(i,j)$ and $J_{2}(i,j)$ are horizontal and vertical
independent random couplings having the same probability distribution, which
reads:

\begin{equation} P(x)=(1-p)\delta(x-J)+p\delta(x-J^{\prime}) \label{b}
\end{equation}

\noindent Also, $p$ is the concentration of impurity bonds and both $J$ and
$J^{\prime}$ are assumed to be positive so that the Hamiltonian favors aligned
spins. Notice that both antiferromagnetic couplings (creating frustration) and
broken bonds ($J^{\prime}=0$) lead to ambiguities in the transfer matrix and
must be excluded in the present treatment.  Let us now consider the calculation
of the partition function of the model under discussion

\begin{equation} Z=\sum \exp(-\frac{H}{T}) \label{c} \end{equation}

\noindent where $H$ is defined in eq.(\ref{a}) and the sum runs over all
$2^{N^{2}}$       possible spin configurations.  The partition function is
known to be represented as the trace of the product of the row-to-row transfer
matrices $\hat{T}_{i}$ \cite{Bax,Kog,EFLS}:

\begin{equation} Z=Tr\prod_{i=1}^{N}\hat{T}_{i} \label{d} \end{equation}

\noindent The Hermitian $2^{N}\times 2^{N}$ matrix $\hat{T}_{i}$ rewritten in
terms of spin variables reads \cite{Bax,Kog,EFLS,Shalaev2}:

\begin{equation} \hat{T}_{i}=\exp \left (
\frac{1}{T}\sum_{j=1}^{N}J_{1}(i,j)\sigma_{3}(j) \sigma_{3}(j+1) \right ) \exp
\left ( \frac{1}{T}\sum_{l=1}^{N}J_{2}^{*}(i,l)\sigma_{1}(l) \right ) \label{e}
\end{equation}

\noindent where $\sigma_{\alpha}$, $\alpha=1,2,3$ are Pauli spin matrices; here
$J_{2}$ and $J_{2}^{*}$ are related by the Kramers-Wannier duality relation
\cite{Bax,Kog,EFLS}:

\begin{equation} \tanh(\frac{J_{2}^{*}}{T})=\exp(-\frac{2J_{2}}{T}) \label{f}
\end{equation}

\noindent In eq.(\ref{e}) we have set an irrelevant factor to unity. Since
 the non-averaged operator $\hat{T}_{i}$ in eq.(\ref{g}) is random, the
representation in eq.(\ref{d}) is in fact inappropriate for computing the
partition function.  In order to get a more convenient starting point for
further calculations we apply the replica trick. We introduce $n$ identical
"replicas" of the original model labeled by the index $\alpha$,
$\alpha=1,\dots,n$ and use the well-known identity for the averaged free
energy:

\begin{equation} \bar{F}=-T\overline{ \ln Z }= -T \lim_{n\rightarrow
0}\frac{1}{n}\overline{(Z^n -1)} \label{g} \end{equation}

\noindent Substituting eq.(\ref{d}) into eq.(\ref{g}) one obtains:

\begin{equation} \overline{F}=-T\lim_{n\rightarrow 0
}\{\overline{Tr\prod_{\alpha=1}^{n}
\prod_{i=1}^{N}\hat{T}_{i}^{\alpha}-1}\}\frac{1}{n} \label{h} \end{equation}

\noindent In contrast to the case of random-site disorder, for the random-bond
problem the two matrices $\hat{T}_{i}^{\alpha}$ and $\hat{T}_{j}^{\beta}$ with
different row indices $ i \neq j $ depend on two different sets of random
coupling constants and commute           to each other for any $\alpha$ and
$\beta$. This allows us to average these two operators independently. After
some algebra one arrives at

\begin{equation} \overline{Z^{n}}=Tr\hat{T}^{N} \label{i} \end{equation}

\noindent where the tranfer matrix $\hat{T}$ of the 2D random-bond IM is given
by \cite {Shalaev2}:

\begin{eqnarray} \hat{T}&&=\overline{\prod_{\alpha=1}^{n}\hat{T}_{i}^{\alpha}}
\nonumber \\ &&=\exp \left \{
\sum_{j=1}^{N}\log[(1-p)\exp(\frac{J}{T}\sum_{\alpha=1}^{n}
\sigma_{3}^{\alpha}(j))
\sigma_{3}^{\alpha}(j+1)+p\exp(\frac{J^{\prime}}{T}\sum_{\alpha=1}^{n}
\sigma_{3}^{\alpha}(j)\sigma_{3}^{\alpha}(j+1))] \right \} {\times} \nonumber
\\ &&{\times}\exp \left \{ \sum_{j=1}^{N}\log \left [ (1-p)\exp \left (
\frac{J^{*}}{T}\sum_{\alpha=1}^{n} \sigma_{1}^{\alpha}(j) \right ) +p\exp
(\frac{J^{*\prime}}{T}\sum_{\alpha=1}^{n}\sigma_{1}^{\alpha}(j)) \right ]
\right \} \label{j} \end{eqnarray}

\noindent Setting $p$ to zero (or $J=J^{\prime}$) one is indeed led to the well
known expression for the T-operator of the pure IM \cite{Bax}:

\begin{equation}
\hat{T}_{PIM}=\exp\{\frac{J}{T}\sum_{j=1}^{N}\sigma_{3}(j)\sigma_{3}(j+1)\}
\exp\{\frac{J^{*}}{T}\sum_{j=1}^{N}\sigma_{1}(j)\} \label{k} \end{equation}

\noindent The T-matrix is known to possess the Kramers-Wannier dual symmetry.
In the language of spin variables this nonlocal mapping reads \cite{Kog,EFLS}:

\begin{equation} \tau_{1}(k)=\sigma_{3}(k)\sigma_{3}(k+1) \qquad
\tau_{2}(k)=i\sigma_{1}(k)\sigma_{3}(k) \qquad
\tau_{3}(k)=\prod_{m<k}\sigma_{1}(m) \label{l} \end{equation}

\noindent where the operators $\tau_{\alpha}(k)$ satisfy the very same algebra
as the Pauli spin matrices $\sigma_{\alpha}(n)$. It is easy to see that if
$p=0,\frac{1}{2},1$ the T-matrix given by eq.(\ref{j}) is invariant under the
dual transformation. The plausible assumption that there is a single critical
point yields the equation for the critical temperature $T_{c}$:

\begin{equation} \exp (-\frac{2J^{\prime}}{T_{c}})=\tanh(\frac{J}{T_{c}})
\label{m} \end{equation}

\noindent Notice that the point $p=\frac{1}{2}$ is not the percolation
threshold, because the coupling constants $J$ and $J^{\prime}$ are assumed to
take nonzero values with the ferromagnetic sign.  Writing $\hat{T}$ in the
exponential form

\begin{equation} \hat{T}=\exp(-\hat{H}) \label{n} \end{equation}

\noindent one obtains the partition function in the following form:

\begin{equation} Z=Tr\exp(-N\hat{H}) \label{o} \end{equation}

\noindent where by definition $\hat{H}$ is just the logarithm of the transfer
matrix $\hat{T}$ (the "quantum" Hamiltonian). In the thermodynamic limit $N
\rightarrow \infty$ the free energy is proportional to the lowest eigenvalue of
the "quantum" Hamiltonian $\hat{H}$ \cite{Kog,EFLS}:

\begin{equation} F=-T\ln Tr\exp(-N\hat{H})\rightarrow NTE_{0} \qquad
\hat{H}|0>=E_{0}|0> \label{p} \end{equation}

\noindent Here $|0 \rangle$ is the ground state of $\hat{H}$ which is assumed
to be non-degenerate. Actually this means that we assume $T>T_c$. From
eq.(\ref{j}) and eq.(\ref{n}) it follows that $\hat{H}$ is not a simple local
operator. A crucial simplification occurs by taking the y-continuum limit with
$a_{y}\rightarrow 0$ (the lattice spacing     along the y-axis). In other
words, after calculating the logarithmic derivative of $\hat{T}$ with respect
to $a_{y}$ and setting $a_{y}$ to zero the "quantum" Hamiltonian takes on the
following simple form (for details see \cite{Shalaev2}):

\begin{eqnarray} \hat{H}&=&\frac{d \ln\hat{T}}{da_{y}}|_{(a_{y}=0)}
=-\sum_{j=1}^{N} \left \{ K_{1}\sigma_{3}^{\alpha}(j)\sigma_{3}^{\alpha}(j+1) +
K_{2}\sum_{\alpha=1}^{n}\sigma_{1}^{\alpha}(j) \right. \nonumber \\ &+& \left.
K_{4}^{\prime}(\sigma_{3}^{\alpha}(j)\sigma_{3}^{\alpha}(j+1))^{2} +
K_{4}^{\prime\prime}(\sum_{\alpha=1}^{n}\sigma_{1}^{\alpha}(j))^{2} \right \}
\label{q} \end{eqnarray}

\noindent The higher-order terms in the spin operators are known to be
irrelevant in the critical region, so that they can be dropped in eq.(\ref{q}).
The replicated Hamiltonian, eq.(\ref{q}), may be converted into the fermionic
one by means of the Jordan-Wigner transformation \cite{Kog,EFLS}:

\begin{eqnarray}
c^{\alpha}(m)&=&\sigma_{-}^{\alpha}(m)\prod_{j=1}^{m-1}\sigma_{1}^
{\alpha}(j)Q^{\alpha} \nonumber\\
c^{\alpha^{+}}(m)&=&\sigma_{+}^{\alpha}(m)\prod_{j=1}^{m-1}\sigma_{1}^{\alpha}
(j)Q^{\alpha} \nonumber\\
\sigma_{\pm}&=&\frac{1}{2}(\sigma_{3}\pm\ i\sigma_{2}), \qquad
Q^{\alpha}=\prod_{\beta=1}^{\alpha-1}\prod_{j=1}^{N}\sigma_{1}^{\beta}(j),
\qquad \alpha=1,\dots,n \label{r} \end{eqnarray}

\noindent where $c^{\alpha}(m)$ and $c^{\alpha^{+}}(m)$ are the standard
annihilation and creation fermionic operators which satisfy the canonical
anticommutation relations:

\begin{equation}
\{c^{\alpha}(m),c^{\beta^{+}}(n)\}=\delta^{\alpha\beta}\delta_{mn}\qquad
\{c^{\alpha}(m),c^{\beta}(n)\}=0 \label{s} \end{equation}

\noindent After making different species anticommute, the Klein factors
$Q^{\alpha}$ drop out of $\hat{H}$.  For each species it is convenient to
introduce a two-component Hermitean Majorana spinor field \cite{ZI}:

\begin{eqnarray}
\psi_{1}^{\alpha}(n)&=&\frac{1}{\sqrt{2a_{x}}}[c^{\alpha}(n)\exp(-i\frac{\pi}
{4}) +c^{\alpha^{+}}(n)\exp(i\frac{\pi}{4})] \nonumber \\
\psi_{2}^{\alpha}(n)&=&\frac{1}{\sqrt{2a_{x}}}[c^{\alpha}(n)\exp(i\frac{\pi}{4})
+c^{\alpha^{+}}(n)\exp(-i\frac{\pi}{4})] \label{t} \end{eqnarray}

\noindent with standard anticommutation rules:

\begin{equation}
\{\psi_{c}^{\alpha}(n),\psi_{b}^{\beta}(m)\}=\frac{1}{a_{x}}\delta^{\alpha\beta}
\delta_{bc}\delta_{mn}\qquad c,b=1,2.  \label{u} \end{equation}

\noindent where $a_{x}$ is the lattice spacing along the x-axis. Using
eq.(\ref{u}) and the following relations

\begin{eqnarray} \sigma_{1}^{\alpha}(n) &=& 2c^{\alpha^{+}}(n)c^{\alpha}(n)-1
\nonumber\\
\sigma_{3}^{\alpha}(n)\sigma_{3}^{\alpha}(n+1)&=&[c^{\alpha^{+}}(n)-c^{\alpha}
(n)][c^{\alpha^{+}}(n+1)+c^{\alpha}(n+1)] \label{v} \end{eqnarray}

\noindent one can easily rewrite the Hamiltonian, eq.(\ref{q}), in terms of
Majorana fermionic fields. Now let us notice that in the vicinity of $T_{c}$
the correlation length $\xi$ goes to infinity and the system ``forgets'' the
discrete nature of the lattice. For that reason we can simplify the Hamiltonian
by taking the continuum limit $a_{x}\rightarrow 0$. Perfoming simple but
cumbersome calculations we arrive at the $O(n)$-symmetric Lagrangian of the
Gross-Neveu model \cite{DD}

\begin{equation} L=\int
d^{2}x[i\bar{\psi_{a}}\hat{\partial}\psi_{a}+m_{0}\bar{\psi_{a}}\psi_{a}
+u_{0}(\bar{\psi_{a}}\psi_{a})^{2}] \label{w} \end{equation}

\noindent where $\gamma_{\mu}=\sigma_{\mu},
\hat{\partial}=\gamma_{\mu}\partial_{\mu}, \mu=1,2,
\bar{\psi}=\psi^{T}\gamma_{0}$ and

\begin{equation} m_{0}\sim K_{1}-K_{2}\sim\tau=\frac{T-T_{c}}{T_{c}}\qquad
u_{0}\sim K_{3} +K_{4} \label{y} \end{equation}

\noindent Here $m_{0}, u_{0}$ are the bare mass of the fermions and their
quartic coupling constant, respectively. Notice, that if $p\ll 1$, $u_{0}\sim
p$. Provided $p=\frac{1}{2}$ and $T=T_{c}$ we have $u_{0}\sim
(J-J^{\prime})^{2}$.

The RG calculations in the one-loop approximation are very simple. In fact, the
$O(n)$-symmetric Gross-Neveu model being infrared-free in the replica limit
$n\rightarrow 0$, the one-loop approximation truly captures the essentials of
the critical behavior of the model under consideration.  The one-loop RG
equations and initial conditions are given by:

\begin{eqnarray} \frac{du}{dt}&=&\beta(u)=-\frac{(n-2)u^{2}}{\pi};\qquad
\frac{d\ln F}{dt}         =-\gamma_{\bar{\psi}\psi}(u)=\frac{(1-n)u}{\pi}
\nonumber\\ u(t=0)&=&u_{0};\qquad F(t=0)=1 \label{z} \end{eqnarray}

\noindent where $u$ is the dimensionless quartic coupling constant, $\beta(u)$
is the Gell-Mann-Low function, $\gamma_{\bar{\psi}\psi}(u)$ is the anomalous
dimension of the composite operator $\bar{\psi}\psi=\epsilon(x)$ (in fact, the
energy density operator), $t=\ln\frac{\Lambda}{m}$, $\Lambda=a^{-1}$ is an
ultraviolet cutoff and $a$ and $m$ are the lattice spacing and renormalized
mass, respectively.  Here $F$ is the following Green's function at zero
external momenta:

\begin{equation} F=\frac{dm}{d\tau}=\int d^{2}x d^{2}y \langle
\bar{\psi}(x)\psi(y)\bar{\psi}(0)\psi(0) \rangle \label{zz} \end{equation}

\noindent The solution of these equations gives the temperature dependence of
the correlation length $\xi$ and specific heat $C$ in the asymptotic region
$t\rightarrow  \infty, n=0$ \cite{DD}:

\begin{eqnarray} u&=&\frac{\pi}{2t};\qquad F\sim
\tau^{-\frac{1}{2}};\nonumber\\ \xi&=&m^{-1}\sim
\tau^{-1}[\ln\frac{1}{\tau}]^{\frac{1}{2}};\nonumber\\ C&\sim&\int dt
F(t)^{2}\sim \ln\ln\frac{1}{\tau} \label{q1} \end{eqnarray}

These results follow from the solution of the one-loop RG equations,
eq.(\ref{z}), but in fact it is worth noticing that they are a direct
consequence of a renormalization statement valid to all orders in perturbation
theory. Consider a version of the field theory, eq.(\ref{w}), in which the
quartic term is decoupled by the introduction of a scalar Hubbard-Stratonovich
field $\phi$:

\begin{equation} L=\int d^2x [ \bar{\psi}_a (i\hat{\partial}+m_0) \psi_a
+\frac{1}{2}\phi^2 +\frac{1}{2} g_0\phi\bar{\psi}_a\psi_a ] \label{g1}
\end{equation}

\noindent with $g_0{\propto}\sqrt{u_0}$. As a consequence of the functional
version of the classical equation of motion \cite{gj1},

\begin{equation} \frac{\delta L}{\delta \phi}=\phi + \frac{1}{2} g_0
\bar{\psi}_a\psi_a =0 \label{g2} \end{equation}

\noindent the vertex parts $\Gamma$ of the correlation functions
$G^{(2,0;1)}_{ab}=\langle \psi_a(x) \psi_b(y) \frac{1}{2}\bar{\psi}_c(z)
\psi_c(z) \rangle$ and $G^{(2,1)}_{ab}=\langle \psi_a(x) \psi_b(y) \phi(z)
\rangle$ are linked by the relationship

\begin{equation} \Gamma^{(2,1)}=-g_0\Gamma^{(2,0;1)}_{\bar{\psi}\psi}
\label{g3} \end{equation}

\noindent where it has been indicated explicitly that the quadratic insertion
refers to the $O_1=\bar{\psi}_a\psi_a$ operator. Imposing the renormalization
conditions, eq.(\ref{g3}) leads to \cite{gj1}

\begin{equation} g_0/g=\sqrt{u_0/u}=Z_{\phi}^{-1/2}Z_{11} \label{g4}
\end{equation}

\noindent where $g$ is the renormalized coupling constant, $Z_{\phi}$ the
$\phi$- field renormalization constant and $Z_{ij}$ is the quadratic-insertion
renormalization matrix for the operators $\{ O_i \} = \{ \bar{\psi}\psi, \phi^2
\}$.  Since for $n=0$ we have $Z_{\phi}=1$, eq.(\ref{g4}) leads to the exact
result $\beta^{(0)}(u)=-2u\gamma^{(0)}_{\bar{\psi}\psi}(u)$ between the
Gell-Mann-Low and the anomalous dimension functions, implying
$2\gamma_1/\beta_2=-1$ for the coefficients of the lowest-order nonzero terms
in the expansion of these functions in $u$ (that is: $\beta(u)=\beta_2
u^2+\cdots$, $\gamma_{\bar{\psi}\psi}(u)=\gamma_1 u+\cdots$). Solving, for
instance, the RG equation for the specific heat function leads to the
remarkable Dotsenko and Dotsenko's result for the leading asymptotic behavior,
when $\tau{\rightarrow}0$

\begin{equation} C{\sim}\int^{\tau} \frac{dx}{x} | \ln x
|^{2\gamma_1/\beta_2}{\sim} \ln \ln \tau \label{g5} \end{equation}

\noindent by virtue of the above $n=0$ exact results. Similar considerations
lead to the announced behavior of the correlation length, $\xi$.

The main conclusion of this Section is that the critical behavior of the 2D
random bond IM is governed by the pure Ising fixed point. It implies that all
critical exponents of the weakly disordered system are the very same as for the
pure model. Randomness gives rise to the self-interaction of the spinor field
which leads to logarithmic corrections to power laws. In the special case
$p=\frac{1}{2}$ duality imposes strong restrictions, in particular it gives the
exact value of the critical temperature $T_{c}$ which is believed to be unique.
At the critical   point the original  lattice model and its continuum version
described by the    Gross-Neveu model Lagrangian become massless, irrespective
of the value of $n$.  We conjecture that there are only two phases divided by
the single critical point given by the self-duality equation, eq.(\ref{m}). It
implies that under this assumption the Griffiths phase shrinks to zero.

\subsection{ Two-spin Correlation Function at Criticality }

In order to complete the calculation of the temperature dependence of other
thermodynamic quantities we have to compute the susceptibility and spontaneous
magnetization near $T_{c}$. For these calculations we need to find the
large-distance asymptotic behavior of the two-spin correlation function at
criticality. The most effective way for calculating different       correlation
functions for the 2D IM is to use bosonization. Below we shall give a brief
description of this procedure, exploiting simple physical arguments.

Before recalling the principles of the bosonization method, however, let us
show how a straight formulation of the problem in terms of pure fermionic
fields leads to some difficulties even in the case of the calculation of the
pure Ising model correlation function's exponent $\eta$ ($=\frac{1}{4}$) at
criticality.  As shown, e.g., by Samuel \cite{gj2}, the two-spin correlation
function can be expressed in the lattice formulation as the partition function
of a defective lattice where along the line $T_{0R}$ of bonds joining the two
sites $(0,0)$ and $(0,R)$ the ``bond strengths''  $\lambda_y{\equiv}\tanh
(J_2/T)$ must be replaced by $\lambda_y^{-1}$. Namely:

\begin{equation} G_y(R)= \langle s_{00}s_{0R} \rangle = \lambda_y^R \langle
\exp \{ -(\lambda_y - \lambda_y^{-1})\sum_{ij{\in}T_{0R}}
y^{\dagger}_{ij}y_{ij+1} \} \rangle \label{g6} \end{equation}

\noindent were the lattice (y-) Grassmann variables $\{ y^{\dagger}_{ij},
y_{ij} \}$ have been introduced \cite{gj2,gj3}. After suitable transformations,
leading to the quadratic term of the effective Grassmann action in eq.(\ref{w})
without replicas, and in the continuum limit, the above eq.(\ref{g6}) reads

\begin{equation} G_y(R,T_c)=\lambda_{yc}^R \langle \exp iT_0 \int_0^R dy
\bar{\psi}(0,y)\psi(0,y) \rangle \label{g7} \end{equation}

\noindent with $T_0=(\lambda_c^{-1} - \lambda_c)/2\lambda_c=\sqrt{2}+1$ at
criticality ($\lambda_c=\sqrt{2}-1$ for the isotropic model). The two-component
Grassmann (or Majorana) field is the same as in eq.(\ref{w}) and is given by
$\psi=a^{-1} ( y^{\dagger} ~~~ y )$. A possible strategy \cite{DD} is now to
evaluate the $R{\rightarrow}\infty$ behavior of $\ln G_y(R,T_c)$ through an
expansion in powers of $T_0$. Use must be made of the propagator
($\hat{x}=x_{\mu}\gamma_{\mu}$)

\begin{equation} S_0(x-x')=\langle \bar{\psi}(x)\psi (x') \rangle _0
=\frac{i}{2\pi}[\hat{x}- \hat{x}']^{-1}f_{\Lambda}(x-x') \label{g8}
\end{equation}

\noindent where $f_{\Lambda}$ is some cutoff function. The typical term in the
expansion for $\ln G(R)$ involves the multiple integral

\begin{eqnarray} &&{\cal I}_{2n}(R)=\int_0^R dy_1 dy_2 \cdots dy_{2n} Tr [
S_0(y_1-y_2)S_0(y_2-y_3) \cdots S_0(y_{2n}-y_1) ] \nonumber \\ &&=a_nR+b_n \ln
R +\cdots \label{g9} \end{eqnarray}

\noindent from which the $R{\rightarrow}\infty$ critical correlator could be
evaluated through

\begin{equation} \ln G(R)=-\sum_{n=1}^{\infty} \frac{(2T_0)^{2n}}{4n}{\cal
I}_{2n}(R) +R \ln \lambda_c \label{g10} \end{equation}

\noindent (the odd-valued power terms vanishing). Taking the (conjectural)
point of view that all terms in $R$ must cancel exactly, the evaluation of the
$\ln R$ terms can proceed \cite{DD} by taking the choice (natural, but leading
to some ambiguities) $f_{\Lambda}=1$ and evaluating every other y-integral
exactly

\begin{eqnarray} &&I_{2n}(R)=\int_0^R \frac{ dy_1 dy_2 \cdots dy_{2n} } {
(y_1-y_2)(y_2-y_3) \cdots (y_{2n}-y_1) } =\int_0^R dy_1 dy_2 \cdots dy_n \times
\nonumber \\ &&\frac{ \ln [(1-R/y_2)/(1-R/y_1)] \ln [(1-R/y_3)/(1-R/y_2)]
\cdots
	\ln [(1-R/y_1)/(1-R/y_n)] }{ (y_1-y_2)(y_2-y_3) \cdots (y_n-y_1) }
\label{g11} \end{eqnarray}

\noindent After a straightforward but laborious reparametrization of the
integral \cite{DD}, we arrive at

\begin{eqnarray} \ln G(R)&=&R \ln \lambda_c - \sum_{n=1}^{\infty} \frac{ (-
T_0^2/\pi^2)^n } {2n} I_{2n}(R) \nonumber \\
I_{2n}(R)&=&\int_{-\infty}^{+\infty} dz_1 dz_2 \cdots dz_{n-1} \frac{ \sum_i
z_i/2 }{ \sinh \sum_i z_i /2 } \prod_i \frac{ z_i/2 } { \sinh z_i /2 }
\int_{\Lambda^{-1}}^R \frac{ Rdx }{ x(R-x) } \nonumber \\ &=&2\theta_n \ln
R\Lambda \label{g12} \end{eqnarray}

\noindent with the $R$-dependence now neatly factorized out and the cutoff
$\Lambda{\sim} a^{-1}$  conveniently reinstated. The coefficient $\theta_n$ is
evaluated through the Fourier representation

\begin{equation} \frac{ z/2 }{ \sinh (z/2) }=\int_{-\infty}^{+\infty} \frac{ dp
}{ 2\pi } F(p) e^{-ipz}, \qquad F(p)=\frac{ \pi^2 }{ \cosh^2 \pi p }
\label{g13} \end{equation}

\noindent leading to $\theta_n=\frac{1}{2\pi} \int_{-\infty}^{+\infty} dp
[F(p)]^n$.  Finally, we get (dropping the $R$-terms)

\begin{equation} \ln G(R)=-\sum_{n=1}^{\infty} \frac{ \theta_n }{n} \left (
\frac{T_0}{\pi} \right )^{2n} \ln R\Lambda = -\eta \ln R\Lambda \label{g14}
\end{equation}

\noindent where

\begin{equation} \eta=\sum_{n=1}^{\infty} \frac{ \theta_n }{n} \left (
\frac{T_0}{n} \right )^2 =\frac{1}{2\pi} \int_{-\infty}^{+\infty} dp
\sum_{n=1}^{\infty} \frac{1}{n} \left [ \frac{T_0^2}{\cosh^2 \pi p} \right ]^n
\label{g15} \end{equation}

\noindent The last sum converges to a logarithm and the $p$-integral can be
evaluated, provided $|T_0|<1$. For $T_0=1$, eq.(\ref{g15}) leads to $\eta=1/4$
\cite{DD}; however, the standard prescription \cite{gj2}  calls for
$T_0=\sqrt{2}+1$ and this leads to a divergence in the summation. Clearly, this
is associated with the use of a uniform cutoff function $f_{\Lambda}=1$, but it
must be stressed that to date no further progress in evaluating the spin-spin
correlator at criticality, using solely the fermionic formalism, can be
reported. The situation is even more delicate when disorder is introduced, thus
the method of the fermionic tail $T_{0R}$ must be abandoned.

Let us now begin discussing bosonization, with the following action:

\begin{equation} L=\int
d^{2}x\{i\bar{\psi}\hat{\partial}\psi+[m_{0}+\tau(x)]\bar{\psi}\psi\}
\label{w1} \end{equation}

\noindent where $\psi$ is a Majorana spinor and $\tau(x)$ is a random Gaussian
field with the following probability distribution:

\begin{equation} P[\tau(x)]\sim\exp\{-\frac{1}{2u_{0}}\int
d^{2}x[\tau(x)]^{2}\} \qquad <\tau(x)\tau(y)>=u_{0}\delta(x-y) \label{e1}
\end{equation}

\noindent In fact, the action eq.(\ref{w1}) describes free fermions moving in
the random potential $\tau(x)$, which in our case is responsible for local
fluctuations of the critical temperature $T_{c}$ in the dilute ferromagnet.
After applying the replica trick and averaging over ``all'' possible
configurations of $\tau(x)$ one gets the very same Gross-Neveu Lagrangian as
given by eq.(\ref{w}).  The representation of the square of the spin-spin
correlation function of the pure 2D IM, that is

\begin{equation} G(x-y)=<\sigma(x)\sigma(y)> \label{r1} \end{equation}

\noindent in terms of the path integral over the real bosonic field $\phi$ of
quantum sine-Gordon model was found by Zuber and Itzykson \cite{ZI} (see also
\cite{FSZ}) and reads:

\begin{eqnarray} G(x-y)^{2}&=&Z^{-1}\frac{1}{2\pi^{2}a^{2}}\int
D\phi\sin(\sqrt{4\pi}\phi(x)) \sin(\sqrt{4\pi}\phi(y))\exp\{-S\}\nonumber\\
S&=&\frac{1}{2}\int d^{2}x\{(\partial_{\mu}\phi)^{2}+\frac{2m_{0}}{\pi a}
\cos(\sqrt{4\pi}\phi)\}\nonumber\\ Z&=&\int D\phi\exp\{-S\} \label{t1}
\end{eqnarray}

\noindent At criticality, $m_{0}=0$, the path integral being Gaussian, the
result of its evaluation is easily seen to be:

\begin{equation} G(x-y)\sim |x-y|^{-\frac{1}{4}} \label{y1} \end{equation}

\noindent The representation for the two-spin correlation function may be
extended to the dilute system by replacing the bare mass $m_{0}\sim\tau$ with
the random one $m_{0}+\tau(x)$ into eq.(\ref{t1}). Of course, in the
inhomogeneous case the non-averaged       $G(x,y)$, being sample dependent,
depends  on $x$ and $y$ separately.  The averaged correlation function
$\overline{G(x-y)}$ at the critical point may be computed (even without using
the replica trick) in two stages:  (i) firstly, the square root of $G(x,y)^{2}$
is formally evaluated by means of expanding it in a power series in $\tau(x)$;
(ii) secondly, the resulting expression is integrated with respect to $\tau(x)$
(for technical details of the calculations see \cite{Shalaev2,ZI1}).  The
conventional RG equation for the renormalized averaged correlation function
reads:

\begin{equation}
\{\mu\frac{\partial}{\partial\mu}+\beta(u)\frac{\partial}{\partial u}
+\eta(u)\}\overline{G_{R}(p,u,\mu)}=0 \label{u1} \end{equation}

\noindent where $\mu$ is a renormalization momentum, $\beta(u)$ is the
beta-function and $\eta(u)$ is defined as follows

\begin{equation} \eta(u)=\beta(u)\frac{d\ln Z_{\sigma}(u)}{du} \label{i1}
\end{equation}

\noindent The spin renormalization constant $Z_{\sigma}(u)$ and the
renormalized correlation function are defined in the standard way:

\begin{equation}
\overline{G(p,u_0,\Lambda)}=Z_{\sigma}(u)\overline{G_{R}(p,u,\mu)} \label{o1}
\end{equation}

\noindent The Kramers-Wannier symmetry was shown to apply in some vanishing
terms         linear in $u$ in the expansions for $\eta(u)$ and $Z_{\sigma}(u)$
\cite{Shalaev2}, that is:

\begin{equation} Z_{\sigma}(u)=1+O(u^{2});\qquad \eta(u)=\frac{7}{4}+O(u^{2})
\label{p1} \end{equation}

\noindent Given $\beta(u)$ and $\eta(u)$ in the one-loop approximation, the
solution of the Ovsyannikov-Callan-Symanzik equation for the correlation
function is quite simple:

\begin{equation} G(p)\sim p^{-\frac{7}{4}};\qquad G(R)\sim R^{-\frac{1}{4}}
\label{a1} \end{equation}

\noindent So, the Fisher critical exponent takes the very same value
$\eta=\frac{1}{4}$ as in the pure model.  Notice that in contrast to higher
moments of the spin correlation function, the first one does not contain the
logarithmic factor due to the above-mentioned dual symmetry \cite{Lud2}. From
this remark it follows that the temperature dependence of the homogeneous
susceptibility and spontaneous magnetization are described by power-law
functions of the correlation length $\xi$ (without logarithmic corrections like
$\ln \xi$)

\begin{eqnarray} \chi&\sim& \xi^{2-\eta}\sim
\tau^{-\frac{7}{4}}[\ln\frac{1}{\tau}]^{ \frac{7}{8}}\nonumber\\ M&\sim&
\xi^{\frac{\eta}{2}}\sim(-\tau)^{\frac{1}{8}}[\ln\frac{1}{(-\tau)}]^{\frac{1}{16}}
\label{s1} \end{eqnarray}

\noindent The equation of state at the critical point may be obtained from the
usual scaling relation:

\begin{equation} H\sim M^{\frac{4+\eta}{\eta}}\sim M^{15} \label{d1}
\end{equation}

\noindent As we predicted, all critical exponents of the quenched disordered
system are identical to those of the pure model, apart from some logarithmic
corrections \cite{Jug}.

\section{THE $N$-COLOR ASHKIN-TELLER MODEL }
\renewcommand{\theequation}{3.\arabic{equation}} \setcounter{equation}{0}

The N-color Ashkin-Teller model (ATM) was introduced by Grest and Widom
\cite{GW} and consists of a system of $N$ 2D Ising models coupled together like
in the conventional 2-color model. The lattice Hamiltonian of the isotropic
$N$-color ATM reads:

\begin{eqnarray} H&=&\sum_{a=1}^{N} H_{I}(s^{a})+J_{4}\sum_{a\neq
b=1}^{N}\sum_{<nn>}\epsilon_{a} \epsilon_{b}\nonumber\\
&=&-\sum_{<nn>}\{J\sum_{a=1}^{N}s_{i}^{a}s_{j}^{a}
+J_{4}[\sum_{a=1}^{N}s_{i}^{a}s_{j}^{a}]^{2}\} \label{q2} \end{eqnarray}

\noindent where $s^{a}=\pm1$, $a=1,\dots,N$, $<>$ indicates that the summation
is over all nearest-neighboring sites, $H_{I}(s^{a})$ is the Hamiltonian of the
pure 2D IM, $\epsilon_{a}=s_{i}^{a}s_{j}^{a}$ is the density energy operator,
and  $J_{4}$ is a coupling constant between the Ising planes.

This model was shown to be the lattice version of a model with hypercubic
anisotropy, describing a set of magnetic and structural phase transitions in
variety of solids \cite{com1,TTMB}.  The corresponding Landau Hamiltonian
reads:

\begin{eqnarray} H&=&\int
d^{2}x\{\frac{1}{2}(\partial_{\mu}\Phi)^{2}+\frac{1}{2}m_{0}^{2}
\Phi^{2}+\frac{1}{8}u_{0}(\Phi^{2})^{2}+\frac{1}{8}v_{0}\sum_{a=1}^{N}
\Phi_{a}^{4}\} \nonumber \\ \Phi^{2}&=&\sum_{a=1}^{N}\Phi_{a}^{2},\qquad
(\partial_{\mu}\Phi)^{2}=\sum_{a=1}^{N}(\partial_{\mu}\Phi_{a})^{2} \label{w2}
\end{eqnarray}

\noindent where $\Phi$ is an $N$-component order parameter,
$m_{0}^{2}\sim\tau$, $u_{0}\sim J_{4}$ and $v_{0}$ are some coupling constants.
In particular, in the replica limit the Hamiltonian, eq.(\ref{w2}), describes
the random-bond IM (for $v_{0}>0, u_{0}<0$). If $v_{0}=0$, a phase transition
in the $O(N)$-symmetric    model with nonzero value of the spontaneous
magnetization is known to be forbidden by the Mermin-Wagner theorem \cite{MW}.
If $v_{0}{\neq}0$ the spontaneous breakdown of the discrete hypercubic symmetry
occurs at $T_{c}>0$. Since the term with $v_{0}$ is strongly relevant, the
perturbation theory expansion with respect to $v_{0}$ is actually hopeless near
$T_{c}$.

By exploiting the operator product expansion (OPE) approach, Grest and Widom
obtained the one-loop $\beta$-function for the quartic coupling constant
$J_4$.  If $J_4<0$ and $N>2$ the phase transition was shown to be continuous
and the critical behavior belonging to the 2D IM universality class \cite{GW}.

The exact solution of the multi-color ATM in the large N-limit was found by
Fradkin \cite{EF}, who developed a rather complicated formalism based on
bosonic fields and showed that a second order phase transition with IM critical
exponents occurs if $J_{4}<0$.  In fact, as was shown by Aharony \cite{AA}, the
model with hypercubic anisotropy, eq.(\ref{w2}), in the large $N$-limit is
equivalent to the IM with equilibrium impurities. Moreover, for the 2D case he
predicted the Ising-type critical behavior with logarithmic corrections.  Such
being the case, one expects that the critical behavior is identical to the IM
critical behavior. Since $\alpha=0$, Fisher's renormalization of the critical
exponents \cite{MF} is inessential and gives rise only to logarithmic factors.
Notice also that in contrast to the pure case the specific heat $C$ is finite
at $T_{c}$.  The exact solution of the 2D IM with equilibrium defects obtained
by Lushnikov \cite {L} many years ago confirms these conclusions.

The effective method for solving the model under discussion, based on a mapping
of the original model, eq.(\ref{q2}), onto the $O(N)$-symmetric Gross-Neveu
model, was suggested by Shankar \cite{Sh2} (see also \cite{Gold,Shalaev3}).  In
order to show this equivalence let us transform the partition function $Z$ by
applying the Hubbard-Stratonovich identity:

\begin{eqnarray} Z&=&\int D\Phi\exp[-H(\Phi)] \nonumber\\ &=&\int D \Phi D
\lambda\exp\{-\int d^{2}x[\frac{1}{2}(\partial_{\mu} \Phi)^{2}
+\frac{1}{2}m_{0}^{2} \Phi^{2}+\frac{1}{8}v_{0}\sum_{a=1}^{N} \Phi_{a}^{4}
+i\lambda(x)\Phi^{2}+\frac{1}{2u_{0}}[\lambda(x)]^{2} ]\}\nonumber\\ &=&\int D
\lambda \exp(-\frac{1}{2u_{0}}\int d^{2}x \lambda^{2}) [Z_{I}[m_{0}^{2}+i
\lambda(x)]]^{N} \label{e2} \end{eqnarray}

\noindent where $\lambda(x)$ is an auxiliary field; $Z_{I}$ is the exact
partition function of the 2D IM which is known to correspond to a path integral
over Grassmann variables (Section II):

\begin{equation} Z_{I}=\int D\bar{\psi} D\psi\exp\{-\int
d^{2}x[i\bar{\psi}\hat{\partial} \psi+\kappa_{0}\bar{\psi}\psi]\} \label{r2}
\end{equation}

 \noindent Now let us replace $\kappa_{0}=m_{0}^{2}$ in eq.(\ref{r2}) by
$m_{0}^{2}+i\lambda(x)$ and substitute eq.(\ref{r2}) into eq.(\ref{e2}). This
replacement is based on the fact that the energy-density operator in the
$\phi^{4}$ theory is $\phi^{2}$ while in 2D fermionic models this is given by
$\bar{\psi}\psi$.  We have \cite{Sh2,Shalaev3}:

\begin{eqnarray} Z&=&\int D\lambda\exp(-\frac{1}{2u_{0}}\int d^{2}x\lambda^{2})
\int \prod_{a=1}^{N}D\bar{\psi_{a}}D\psi_{a}{\times}\nonumber\\ &&\exp\{-\int
d^{2}x[i\bar{\psi_{a}}\hat{\partial}\psi_{a}
+(m_{0}^{2}+i\lambda(x))\bar{\psi_{a}}\psi_{a}]\}\nonumber\\ &=&\int
\prod_{a=1}^{n} D\bar{\psi_{a}} D\psi_{a}\exp(-S_{GN}) \label{t2}
\end{eqnarray}

\noindent where $S_{GN}$ is the Gross-Neveu action, given by eq.(\ref{w}). In
going from eq.(\ref{w2}) to eq.(\ref{t2}) it is assumed that $u_{0}$ has been
rescaled in the following way: $u_{0}\rightarrow u_{0}^{'}=u_{0}a^{-2}$ so as
to make $u_{0}$ dimensionless (the prime will be ignored hereafter).  We see
that the discrete hypercubic symmetry of the N-color ATM evolves into the
continuous $O(N)$ symmetry, hidden    when the system approaches the critical
point.

The one-loop RG equations for the N-color ATM have been already obtained in
Section II, these being  eq.(\ref{z})  where we must set $n=N$. Solving these
equations gives the temperature dependence of the correlation length and
specific heat in the vicinity of the critical point \cite{Shalaev3}:

\begin{eqnarray} \xi&\sim&
\tau^{-1}[\ln(\frac{1}{\tau})]^{\frac{N-1}{N-2}}\nonumber\\
C&\sim&[\ln(\frac{1}{\tau})]^{\frac{N}{2-N}} \label{b2} \end{eqnarray}

\noindent As for the calculation of the correlation function, one can apply the
procedure described in Section II. Like for the random IM case, the term linear
in $u$ for $\eta(u)$ and $Z_{\sigma}(u)$ vanishes due to the Kramers- Wannier
symmetry. This implies the anomalous dimension of the spin $s_{\alpha}$ to be
equal to $\frac{1}{4}$. We get:

\begin{eqnarray} G(R)&\sim& R^{-\frac{1}{4}}\nonumber\\
\chi&\sim&\tau^{-\frac{7}{4}}[\ln\frac{1}{\tau}]^{\frac{7(N-1)}{4(N-2)}}
\nonumber\\
M&\sim&(-\tau)^{\frac{1}{8}}[\ln\frac{1}{(-\tau)}]^{\frac{(N-1)}{8(N-2)}}
\nonumber\\ H&\sim& M^{15} \label{y2} \end{eqnarray}

\noindent Notice that these results are valid only for $N>2,  J_{4}<0$. If
$J_{4}>0$, the discrete $\gamma_{5}$ symmetry $\psi\rightarrow\gamma_{5}\psi,
\bar{\psi} \psi\rightarrow-\bar{\psi}\psi$ is spontaneously broken. From the
$\gamma_{5}$-symmetry breaking it follows that $\langle \bar{\psi}\psi \rangle
\neq 0$. It means that we have a finite correlation length, or in other words,
a first-order phase transition \cite{GW,Sh2}.  So, equations (\ref{y2})
reproduce the well-known results for some particular cases: $N=0,1,\infty$
corresponding to the random-bond IM problem, Onsager problem,  and IM with
equilibrium impurities, respectively.

The symmetric eight-vertex model (or Baxter model) is known to be isomorphic to
the $N=2$ color ATM in the vicinity of the critical line.  The phase diagram of
the 2-color ATM contains the ferromagnetic phase transition line beginning from
the IM critical point and ending at the point corresponding to the 4-state
Potts model. Along this line the model exhibits weakly-universal critical
behavior, with the critical exponents continuously varying. For instance, the
critical exponent $\alpha$ changes continuously from $\alpha=0$ (IM) to
$\alpha=\frac{2}{3}$ (4-state Potts model \cite{Bax}).  The above results
obviously show the special nature of the $N=2$ situation, due to the factor
$\frac{1}{N-2}$. In this case the system under discussion is described by the
$O(2)$-symmetric Gross-Neveu model, or, equivalently, by the massive Thirring
model with the $\beta$-function being equal to zero identically and presenting
nonuniversal critical exponents \cite{Sh2}.  Since the $N=3$ color ATM is
equivalent to the $O(3)$-symmetric Gross-Neveu model which is known to be
supersymmetric \cite{EW}, this model should possess a hidden supersymmetry (see
for details \cite{Sh2,Gold}).

Notice that in contrast to the 2D case, the critical behavior of the $N$-color
ATM in $4-\epsilon$ dimensions ($0<\epsilon\leq 2$) is governed by either the
Gaussian, or the cubic fixed point and never by the IM fixed point. The type of
critical behavior crucially depends on the order parameter component number
$N$. If $N>N_{c}(\epsilon)$, the RG flow arrives at the cubic fixed point; in
the opposite case, $N<N_{c}(\epsilon)$, the Heisenberg (isotropic) fixed point
is stable. Here $N_{c}(\epsilon)$ is the critical dimensionality of the order
parameter, its expansion in powers of $\epsilon$ being \cite{DJW}:

\begin{equation}
N_{c}(\epsilon)=4-2\epsilon-(\frac{5}{2}\zeta(3)-\frac{5}{12})\epsilon^{2}
+O(\epsilon^{3}) \label{u2} \end{equation}

\noindent where $\zeta(3)=1.2020528$ is the Riemann zeta function,
$N_{c}(1)\cong 2.9$ \cite{AIS}. If $\epsilon\rightarrow 2, N_{c}$ decreases and
all the cubic fixed points approach the IM fixed point, merging at
$\epsilon=2$, irrespectively of the value of $N$ \cite{Shalaev3}.

\section{THE GENERALIZED ASHKIN-TELLER MODEL WITH RANDOMNESS }
\renewcommand{\theequation}{4.\arabic{equation}} \setcounter{equation}{0}

Now we extend our study of the $N$-color ATM to two interacting $M$- and
$N$-color quenched disordered Ashkin- Teller models, giving rise to a
generalized Askin-Teller model (GATM). The Landau Hamiltonian is given by

\begin{eqnarray} H&=&\int d^{2}x \left \{ \frac{1}{2}(\partial_{\mu}\Phi)^{2}
+\frac{1}{2} \left [ m_{0}^{2} +\tau_{1}(x) \right ] \Phi_{a}^{2}+\frac{1}{2}
\left [ m_{0}^{2}+\tau_{2}(x) \right ] \Phi_{c}^{2} \right.  \nonumber\\ &+&
\frac{1}{8}u_{1}(\Phi_{a}^{2})^{2}+\frac{1}{8}u_{2}(\Phi_{c}^{2})^{2}
+\frac{1}{8}w_{0}\Phi_{a}^{2}\Phi_{c}^{2} \nonumber\\ &+& \left.
\frac{1}{8}v_{1}\sum_{a=1}^{N}\Phi_{a}^{4}
+\frac{1}{8}v_{2}\sum_{c=N+1}^{N+M}\Phi_{c}^{4} \right \} \label{q3}
\end{eqnarray}

\noindent where $\Phi_{k}, k=1,...,M+N$ is an $(M+N)$-component order
parameter, $a=1,...,N,c=N+1,...,N+M$, $ m_{0}^{2}\sim \tau$,
$v_{\mu},u_{\nu}>0$ and $\mu, \nu=1,2$.  Summation over indices in the
quadratic operators is understood.  We may study two types of impurities: (i)
uncorrelated impurities, and (ii) correlated ones. In these cases the two-point
correlators for the independent random Gaussian fields $\tau_{\mu}$ read:

\begin{equation} (i) \quad \langle \tau_{\mu}(x)\tau_{\nu}(y) \rangle
=z_{\mu}\delta_{\mu\nu}\delta(x-y), \qquad (ii) \quad \langle
\tau_{\mu}(x)\tau_{\nu}(y) \rangle =z_{0}\delta(x-y) \label{w3} \end{equation}

\noindent In fact we study some multicritical point in the model under
discussion, eq.(\ref{q3}), since it has been assumed that
$m_{10}=m_{20}=m_{0}\sim\tau$.  This model in $4-\epsilon$ dimensions (without
disorder) was initially studied by Bruce and Aharony \cite{ABAA} and by
Pokrovskii, Lyuksyutov and Khmelnitskii (without cubic anisotropy) \cite{PKL}
and then in numerous other papers \cite{com1}.  By applying the replica trick
and the ``fermionization'' method described in Section III, one arrives at the
following effective fermionic action involving several types of quartic
fermionic interactions:

\begin{eqnarray} H&=&\int d^{2}x \left \{
i\bar{\Psi_{k}^{\alpha}}\hat{\partial}
\Psi_{k}^{\alpha}+m_{0}\bar{\Psi_{k}^{\alpha}}\Psi_{k}^{\alpha}
+u_{1}\bar{\Psi_{a}^{\alpha}}
\Psi_{a}^{\alpha}\bar{\Psi_{b}^{\alpha}}\Psi_{b}^{\alpha} +
u_{2}\bar{\Psi_{c}^{\alpha}}\Psi_{c}^{\alpha}\bar{\Psi_{d}^{\alpha}}
\Psi_{d}^{\alpha}+ w_{0}\bar{\Psi_{a}^{\alpha}}\Psi_{a}^{\alpha}
\bar{\Psi_{c}^{\alpha}} \Psi_{c}^{\alpha} \right. \nonumber \\ &+& \left.
z_{1}\bar{\Psi_{a}^{\alpha}}\Psi_{a}^{\alpha}\bar{\Psi_{b}^
{\beta}}\Psi_{b}^{\beta} +
z_{2}\bar{\Psi_{c}^{\alpha}}\Psi_{c}^{\alpha}\bar{\Psi_{d}^{\beta}}\Psi_{d}^
{\beta}+ r_{0} \bar{\Psi_{a}^{\alpha}}\Psi_{a}^{\alpha}\bar{\Psi_{c}^{\beta}}
\Psi_{c}^{\beta} \right \} \label{e3} \end{eqnarray}

\noindent where $\Psi_{k}^{\alpha}$ is a (real) Majorana fermionic field,
$\alpha, \beta=1,...,n\rightarrow 0$ are replica indices, $a,b=1,...,N$, and
$c,d=N+1,...,N+M$.  Naively one expects the appearance of two impurity quartic
fermionic couplings in the replicated Hamiltonian, eq.(\ref{e3}). Instead, we
have one additional four-fermion vertex $r_{0}$ (absent in the bare action).
This counterterm arises in the course of the renormalization procedure and
provides  the closedness of the operator algebra. In some sense the appearance
of this term means violating the Harris criterion. The latter is indeed
essentially based on the assumption of the existence of only one operator
responsible for the impurity-induced interaction of the order parameter
fluctuations.

The one-loop RG equations for the six coupling constants $u_{\mu}, v_{\nu}, r$
and $w$ are given by (for $n=0$):

\begin{eqnarray} \frac{du_{1}}{dt}&=&-(N-2)u_{1}^{2}-2z_{1}u_{1}-Mw^{2}
\nonumber \\ \frac{du_{2}}{dt}&=&-(M-2)u_{2}^{2}-2z_{2}u_{2}-Nw^{2} \nonumber
\\ \frac{dw}{dt}&=&-w[(N-1)u_{1}+(M-1)u_{2}+z_{1}+z_{2}] \nonumber \\
\frac{dz_{1}}{dt}&=&-2z_{1}[z_{1}+(N-1)u_{1}+2Mr] \nonumber \\
\frac{dz_{2}}{dt}&=&-2z_{2}[z_{2}+(M-1)u_{2}+2Nr] \nonumber \\
\frac{dr}{dt}&=&-r[(N-1)u_{1}+(M-1)u_{2}+z_{1}+z_{2}]-w[Nz_{1}+Mz_{2}]
\label{r3} \end{eqnarray}

\noindent The initial conditions for both (i) uncorrelated and (ii) correlated
impurities are as follows:

\begin{equation} (i)~~~~ z_{1}(0)=z_{2}(0)=z_{0},~~ r(0)=0, \quad (ii)~~~~
z_{1}(0)=z_{2}(0)=2r(0) =z_{0} \label{y3} \end{equation}

\noindent It is easy to see that if one sets $M=N=1$ (random-bond Baxter or,
equivalently, 2-color ATM) one arrives at the RG equations first obtained by
Dotsenko and Dotsenko \cite{DD1}. In this case the coupling constants $u_{\mu}$
decouple from the others; moreover, instead of two couplings we have only one
coupling constant $z=z_{1}=z_{2}$. It was shown that in almost all cases even
weak disorder would drastically change the critical behavior of the 2-color ATM
from a nonuniversal behavior to the Ising-type one, modified by some
logarithmic corrections (see also \cite{Shalaev2}). Even though the critical
exponent  $\alpha$ of the pure model is negative for $w_{0}<0$, for
uncorrelated defects we find that the critical behavior of the model under
discussion is changed by the emergence of the new scaling field $r$.  In the
case of correlated defects with $w_{0}<0$ the critical behavior of the random
model was shown \cite{DD1} to be still nonuniversal with critical exponents
$\alpha$ and $\nu$ depending on both $w_{0}$ and on the concentration of
impurities $(r_{0})$. This is the only exceptional case in which we would have
nonuniversal critical behavior for a  disordered system.  In all cases the
two-spin correlation function was shown to have, however, the same large
distance behavior as for the 2D IM \cite{Shalaev2}.

Now let us consider two interacting $N$- and $M$-color ATM without randomness
($z_{\mu}=r_{0}=0$). There are several different types of asymptotic behavior
of the coupling constants $u_{\mu}(t), w(t)$, but there is only one stable
solution exhibiting infrared-free behavior. That is given by:

\begin{equation} u_{1}(t)=\frac{1}{(N-2)t},\qquad
u_{2}(t)=\frac{1}{(M-2)t},\qquad w(t)=O(\frac{1}{t \ln t}),\qquad
t\rightarrow\infty \label{u3} \end{equation}

\noindent As a result, the original model decouples into two independent $N$-
and $M$-    color models as described in Section III. Thus, the hidden symmetry
of the model near the critical point is the continuous $O(N)\times O(M)$ group.
There also exists a solution of the RG equations given by

\begin{equation} u_{1}(t)=u_{2}(t)=\pm
\frac{1}{2}w(t)=\frac{1}{(N+M-2)t},\qquad t\rightarrow \infty \label{i3}
\end{equation}

\noindent corresponding to the higher symmetry $O(M+N)$ being explicitly broken
in the original Landau Hamiltonian, eq.(\ref{q3}). This is shown to be
unstable.  For instance, provided $N=2, M=1$ (or viceversa) and
$u_{1}(0)=u_{2}(0) =\frac{1}{2} w(0)$, we would obtain the supersymmetric
asymptotic solution of eq.(\ref{i3}). Were these conditions to be broken, i.e.
were the supersymmetry explicitly broken, this would not be restored in the
infrared limit \cite{Gold}.  Notice that our model without cubic anisotropy and
randomness was shown to exhibit this enhanced asymptotic symmetry in
$(4-\epsilon)$-dimensional space, provided $M+N<4$ \cite{PKL}.

One may expect that, due to the critical decoupling of two multi-color ATM into
two independent models, the quenched disorder does not affect the critical
behavior of the system, eq.(\ref{q3}). This is because if $N,M>2$ the specific
heat is finite at criticality (eq.(\ref{t2})) and randomness is irrelevant in
accordance with the Harris criterion. As was explained above this reasonable
assumption should be checked in view of the obvious breakdown of the Harris
criterion due to the appearance of the additional scaling field $r$. The answer
is that this is indeed the case. In fact, it is easy to check that the solution
given by eq.(\ref{u3}) and describing pure models is stable despite the
presence of three disorder couplings.

Thus, from our RG calculations it follows that, in contrast to a 2D IM with
random bonds,  weak quenched disorder here is irrelevant near $T_{c}$.
Moreover, in the critical region the decoupling of two interacting multi-color
ATM was found to occur even in the presence of quenched disorder. The
temperature dependence of the main thermodynamic quantities near the critical
point, the two-spin correlation function and equation of the state at
criticality of the model under consideration are given by eq.(\ref{t2}) and
eq.(\ref{y2}).

\section{ THE WEAKLY-DISORDERED MINIMAL CONFORMAL FIELD THEORY MODELS }
\renewcommand{\theequation}{5.\arabic{equation}} \setcounter{equation}{0}

The critical behavior of the minimal models of conformal field theory with
$c<1$ and as perturbed by a small amount of impurities is far from being solved
and therefore is of considerable interest. In accordance with the Harris
criterion, weak quenched disorder is expected to be strongly relevant near
criticality since the critical exponent $\alpha$ of these models is always
positive and given by $\alpha=\frac{2(m-3)}{3(m-1)}$, with $m=3,4,\dots $. In
particular, for the 3- and 4-state Potts model we have $\alpha=\frac{1}{3}$,
$(m=5)$, and $\alpha=\frac{2}{3}$, $(m=\infty)$, respectively.

The first results in this field were obtained in some pioneering papers by
Ludwig \cite{Lud3} by and Dotsenko, Picco and Pujol \cite{DPP}. They succeeded
in developing a powerful approach closely connected with the formalism
exploited in the previous Sections for describing the multi-color ATM.  These
authors suggested a special kind of $\epsilon$-expansion for computing the
critical exponents, where now $\epsilon=c-\frac{1}{2}$.  Here $c$ is the
central charge of the minimal models without randomness and $\frac{1}{2}$ is
the conformal anomaly of the 2D IM. The main result of their considerations is
that the $\beta(u)$ and $\gamma_{\bar{\psi}\psi}(u)$ functions coincide with
the corresponding functions for the $O(N)$-symmetric Gross-Neveu model obtained
in the framework of the minimal substraction scheme combined with dimensional
regularization.  The distinguishing feature of this scheme is that these
functions do not depend on $\epsilon$ except for the first term in the
$\beta$-function.  Thus, there is a clear possibility to apply the results of
multi-loop RG calculations for the Gross-Neveu model in order to compute the
critical exponents of random minimal models.  At the present time we have the
5-loop expressions for the $\beta(u)$-function and anomalous-dimension
functions $\gamma_{\psi}(u)$ and $\gamma_{\bar{\psi}\psi}(u)$ of the fermionic
field $\psi$ and composite operator $\bar{\psi}\psi$, respectively, and as
obtained in \cite{ANV}. Unfortunately, these expressions contain a few unknown
coefficients in the 4- and 5-loop terms. As for the anomalous dimension of the
spin variable $\eta(u)$, this function was obtained in \cite{DPP} in a 3-loop
approximation. Notice that according to the conformal field theory
classification, the spin variable corresponds to the operator $\Phi_{m,m-1}$,
whilst $\Phi_{1,2}=\epsilon(x)$ is the energy density operator. Thus, one may
use the RG functions obtained only in the 3-loop approximation for the
calculation of the critical exponents.

Let us now compute the critical exponents of the correlation length and
specific heat of the random minimal models in the 3-loop approximation. The
corresponding expressions for the $\beta$-function and temperature critical
exponent function are given by \cite{ANV}:

\begin{eqnarray} \beta(u)&=&2\epsilon
u-2(N-2)u^{2}+4(N-2)u^{3}+2(N-2)(N-7)u^{4} \nonumber \\
\gamma_{\bar{\psi}\psi}(u)&=&2(N-1)u-2(N-1)u^{2}-2(N-1)(2N-3)u^{3} \nonumber \\
\epsilon&=&\frac{3-m}{2m}, \qquad m=3,4,...  \label{q4} \end{eqnarray}

\noindent Here $N$ is the number of planes (colors), coupled to each other in
the usual way like in the N-color ATM (eq.(\ref{q2})). The critical behavior of
the multi-color minimal models is governed by the nontrivial fixed point of
eq.(\ref{q4}). From this equation it follows that:

\begin{eqnarray}
\frac{1}{\nu}&=&\frac{1}{\nu_{0}}+\gamma_{\bar{\psi}\psi}(u^{*})
 =\frac{1}{\nu_{0}}+2(N-1)[\frac{\epsilon}
{N-2}+\frac{\epsilon^{2}}{(N-2)^{2}}-\frac{N\epsilon^{3}}{(N-2)^{3}}] \nonumber
\\ \nu_{0}&=&\frac{2m}{3(m-1)} \label{w4} \end{eqnarray}

\noindent where $\epsilon$ takes on the discrete values defined in
eq.(\ref{q4}) and $\nu_{0}$ is the critical exponent of the correlation length
of the pure model.

To check the self-consistency of eq.(\ref{q4}) let us consider the limit
$N\rightarrow\infty$, describing the system with equilibrium impurities. The
result is  easily seen to be:

\begin{equation} \nu_{imp}=\frac{2m}{m+3}=\frac{\nu_{0}}{1-\alpha_{0}}
\label{u4} \end{equation}

\noindent
From the expression for the anomalous dimension
of the order parameter $\eta(u)$ obtained in \cite{DPP} it follows that the
Fisher critical exponent $\eta_{0}$ is unchanged in this limit,
$\eta_{imp}=\eta_{0}$, where for the q-state Potts $\eta_{0}$ is given by
\cite{DF}:

\begin{equation} \eta_{0}=\frac{(m+3)(m-1)}{4m(m+1)}, \qquad
q=4\cos^{2}\frac{\pi}{m+1},\qquad m=2,3,5,\infty \label{i4} \end{equation}

\noindent As was expected, we have obtained the duly renormalized critical
exponent of the correlation length and an unchanged value of the order
parameter anomalous dimension, in agreement with the predictions of the general
theory \cite{MF}.  As expected, in the trivial case $N=1$ one gets the critical
exponents of pure systems.

The critical exponent values $\nu_{r}$ for random models can be easily obtained
from eq.(\ref{w4}) by expanding $\nu_{r}$ in powers of $\epsilon$ and setting
$N=0$. The results of these calculations are presented in Table I.  The most
striking feature of the above expansions is that they look like rapidly
convergent series, even in the case of the 4-state Potts model.  As a matter of
fact, this is not so surprising, because in the Thirring model ($N=2$) the
anomalous dimensions $\gamma_{\bar{\psi}\psi}(u)$ and $\gamma_{\psi}(u)$ are
known to be some geometric progressions in $u$.  Notice also that if we set
$N=2$ in eq.(\ref{q4}) we do not obtain these progressions \cite{ANV}. The
reasons why the $O(2)$-symmetric Gross-Neveu model in the minimal substraction
scheme is not completely equivalent to the Thirring model have not been
completely understood as yet \cite{ANV}.  It is also important that all values
of $\nu_{r}$ for any arbitrary integer $m$ slightly exceed unity, so that all
values of $\alpha_{r}$ are negative in full agreement with the Harris
criterion.

As was mentioned above, the 3-loop results concerning the critical exponent
$\eta(u)$ were obtained in \cite{DPP}. It turns out that the numerical values
of $\eta$ are in the close vicinity of the Ising model value, $\eta=0.25$.  The
reasons for that are as follows. The 1-loop approximation term which is
expected to give rise to the main contribution to the deviation of $\eta$ from
$\eta_{0}$ vanishes due to the Kramers-Wannier dual symmetry.  The 2-loop
correction was shown to be proportional to $\epsilon u^{2}$, therefore the
deviations of $\eta$ from the pure values are proportional to $\epsilon^{3}$
\cite{DPP}.

Thus, the numerical values of critical exponents of the weakly-disordered
minimal models are very close to the critical exponents of the 2D IM. This has
clear implications for the 3- and 4-state Potts models with random bonds, as
shown by Table I. From a numerical point of view one might be tempted to
conclude that all these models are described by the 2D IM fixed point
\cite{conj}. We see here that this ``superuniversality'' is only approximate,
though accidentally verified to a high degree of accuracy.

\section{ CONCLUSIONS } \renewcommand{\theequation}{6.\arabic{equation}}
\setcounter{equation}{0}

It has been shown that the critical behavior of a good number of 2D anisotropic
systems controlled by the IM fixed point is stable in the presence of weak
quenched disorder. This statement was found to hold quite generally for the 2D
IM, multi-color ATM, and some of its generalizations for which randomness is
marginally relevant. In the case of the 2-color ATM, or Baxter model, disorder
drastically changes the nonuniversal critical behavior inherent in this model
over to the Ising-type critical behavior.  Although some of these models
exhibit a breakdown of the Harris criterion, this does not affect, in general,
the stability of the IM fixed point.  It is commonly believed that the type of
randomness (random-bond, or site-disorder) does not play a role near $T_c$,
despite the fact that random-site disorder has not been studied in great detail
as yet.

On the numerical side, Monte Carlo simulation results are in good agreement
with the analytical results based on the RG calculations \cite{WS}. For
instance, the high-accuracy MC simulation results for a $1024\times 1024$ Ising
lattice with ferromagnetic impurity bonds, recently obtained by Schur and
Talapov \cite{WS}, show that the exponent of the two-spin correlation function
at criticality is numerically very close to that for the pure model. On the
other hand, numerical results obtained by Domany and Wiseman \cite{conj} do
somewhat contradict the theoretical predictions. These results, for a
$256\times256$ lattice, favor a log-type behavior of the specific heat near
$T_{c}$ for the disordered 2-color ATM and 4-state Potts models, and a
double-log behavior of the specific heat for the random-bond IM. As was
established by Dotsenko and Dotsenko \cite{DD1}, the specific heat of the
random 2-color ATM should exhibit the double-log divergence at the critical
point.

The critical behavior of the random 4-state Potts model was shown to be
described by a new fixed point which does not coincide with the IM one
\cite{DPP,Lud3}. The conjecture made in \cite{Shalaev2} that the perturbation
theory expansion around the free fermion theory appropriate for the Ising model
is valid till the end point of the ferromagnetic phase transition line
(describing the 4-state Potts model) is actually incorrect. Exact results for
the repulsion sector of the sine-Gordon theory obtained by means of the Bethe
ansatz in \cite{VEK} show that the perturbation theory around $\beta^{2}=4\pi$
(free fermions) diverges at $\beta^{2}=\frac{16}{3\pi}$ (see also \cite{NITO}),
i.e. it has a finite radius of convergence, and cannot be continued to
$\beta^{2}=8\pi$.

The critical exponents corresponding to the disordered 4-state Potts fixed
point slightly differ from the IM ones. The numerical results are believed to
be sensitive to that difference. It is interesting to compare the estimate for
the critical exponent $\nu$ of the 4-state Potts model based on the 3-loop
approximation with known numerical results. From Table I it follows that
$\nu=1.081$. Novotny and Landau \cite{NL} obtained for the Baxter-Wu model
(equivalent to the 4-state Potts model) the following value: $\nu=1.00(7)$.
The result of Andelman and Berker \cite{AB} is given by: $\nu=1.19$.  Finally,
the recent result obtained by Schwenger, Budde, Voges, and Pfn\"ur \cite{SBVP}
is as follows: $\nu=1.03(8)$.

We end this section by giving a remark that logarithmic corrections to the
power-law dependence and corrections to scaling may give rise to a dependence
of effective critical exponents on the concentration of defects as observed in
some numerical experiments \cite{RK}.

\begin{center} ACKNOWLEDGEMENTS \end{center}

The authors are most grateful to the National Institute for Nuclear Research
(INFN) in Pavia and to the Interdisciplinary Laboratory of the International
School for Advanced Studies in Trieste, where a considerable part of this work
was carried out, for support, warm hospitality and the use of its facilities.
One of the authors (BNS) has benefitted from numerous discussions with the
participants of the Workshop "Phase Transitions in Dilute Systems" in Bad
Honnef (July, 1995).  He is also grateful to Vik.S.Dotsenko and P.Pujol for
making their preprints available prior to publication.

\begin{table} \begin{tabular}{rccccccc} ${\rm model}$ & $m$ & $\epsilon$ &
$\nu_{o}$ & $\nu_{1}$ & $\nu_{2}$ & $\nu_{3}$ & $\nu_{r}$ \\ \tableline
 TIM  & 4 & -0.125 & 0.889 & 0.099 & 0.017 & 0.003 & 1.008 \\ 3-PM & 5 & -0.2 &
 0.833 & 0.139 & 0.038 & 0.008 & 1.018 \\ TPM  & 6 & -0.25 & 0.8 & 0.16 & 0.052
 & 0.014 & 1.026 \\ 4-PM & $\infty$ & -0.5 & 0.667 & 0.222 & 0.13 & 0.062 &
 1.081 \\ \end{tabular} \caption{ Critical correlation length exponent $\nu$
for random minimal models: TIM (Tricritical Ising Model), 3-PM (3-state Potts
Model), TPM (Tricritical Potts Model), and 4-PM (4-state Potts Model). $m$
denotes the minimal model, $\epsilon=(3-m)/2m$, $\nu_{0}$ is the homogeneous
exponent, $\nu_{r}=\nu_{0}+\nu_{1}+\nu_{2}+\nu_{3}$ the random one and
$\nu_{n}$ denotes the $n$-loop contribution to $\nu_{r}$ } \end{table}

\end{document}